\documentclass[showpacs,preprintnumbers,amsmath,amssymb,prb,twocolumn]{revtex4}
\usepackage[dvips]{graphics,color}
\usepackage{dcolumn}
\begin{document}
\title{Gilbert Damping in Conducting Ferromagnets II:\\
Model Tests of the Torque-Correlation Formula}
\author{Ion Garate}
\author{Allan MacDonald}
\affiliation{Department of Physics, The University of Texas at Austin, Austin TX 78712}
\date{\today}

\begin{abstract}
We report on a study of Gilbert damping due to particle-hole pair excitations in conducting ferromagnets.
We focus on a toy two-band model and on a four-band spherical model which provides an approximate 
description of ferromagnetic (Ga,Mn)As.  These models are sufficiently simple that 
disorder-ladder-sum vertex corrections to the long-wavelength spin-spin response function can be 
summed to all orders.  An important objective of this study is to assess the reliability of 
practical approximate expressions which can be combined with electronic structure 
calculations to estimate Gilbert damping in more complex systems.
\end{abstract}

\maketitle

\section{Introduction}

The key role of the Gilbert parameter $\alpha_{G}$ in current-driven\cite{ralphstiles} and precessional\cite{heinrich} magnetization reversal 
has led to a renewed interest in this important 
magnetic material parameter.  The theoretical foundations which relate Gilbert damping to the transverse spin-spin response function of the 
ferromagnet have been in place for some time\cite{prange,oldkambersky}.  It has nevertheless been
difficult to predict trends as a function of temperature and across materials systems, 
partly because damping depends on the strength and nature of the disorder in a manner 
that requires a more detailed characterization than is normally
available. Two groups have recently\cite{realistic} reported successful applications to transition metal
ferromagets of the {\em torque-correlation} 
formula\cite{oldkambersky,realistic,alphaI} for $\alpha_{G}$.  This formula has the important advantage 
that its application requires knowledge only of the band structure, including its  
spin-orbit coupling, and of Bloch state lifetimes.  The torque-correlation formula is physically 
transparent and can be applied with relative ease in combination with modern 
spin-density-functional-theory\cite{sdft} (SDFT) electronic structure calculations.  In this paper we compare the predictions of the 
torque correlation formula with Kubo-formula self-consistent-Born-approximation
results for two different relatively simple model systems, an artificial two-band model
of a ferromagnet with Rashba spin-orbit interactions and a four-band model which 
captures the essential physics of (III,Mn)V ferromagnetic semiconductors\cite{gamnas}.
The self-consistent Born approximation theory for $\alpha_{G}$ requires that 
ladder-diagram vertex corrections be included in the transverse spin-spin response function. 
Since the Born approximation is exact for weak scattering, we can use this comparison
to assess the reliability of the simpler and more practical torque-correlation formula.
We conclude that the torque-correlation formula is accurate when the Gilbert damping is 
dominated by intra-band excitations of the transition metal Fermi sea, but that it can be inaccurate when 
it is dominated by inter-band excitations.  

Our paper is organized as follows.  In Section II we explain how we evaluate the transverse 
spin-spin response function for simple model ferromagnets.  Section III 
discusses our result for the two-band Rashba model while Section IV summarizes our findings for the 
four-band (III,Mn)V model.  We conclude in Section V with a summary of our results and recommended 
{\em best practices} for the use of the torque-correlation formula.

\section{Gilbert Damping and Transverse Spin Response Function} 

\subsection{Realistic SDFT {\em vs.} s-d and p-d models}

We view the two-band $s-d$ and four band $p-d$ models studied in this paper as 
toy models which capture the essential features of 
metallic magnetism in systems that are, at least in principle\cite{phenomenological},
more realistically described using SDFT.  The $s-d$ and $p-d$ models 
correspond to the limit of {\em ab initio} SDFT in which i) the 
majority spin $d$-bands are completely full and the minority 
spin $d$-bands completely empty, ii) hybridization between  
$s$ or $p$ and $d$-bands is relatively weak, and iii) there is exchange
coupling between $d$ and $s$ or $p$ moments.  In a recent paper we have 
proposed the following expression for the Gilbert-damping contribution from
particle-hole excitations in SDFT bands:
\begin{equation} 
\label{eq:alpha_def}
\alpha_{G} =  \frac{1}{S_0} \; \partial_{\omega} {\rm Im}[\tilde{\chi}^{QP}_{x,x}] 
\end{equation} 
where $\tilde{\chi}^{QP}_{x,x}$ is a response-function which describes how the 
quasiparticle bands change in response to a spatially smooth variation in 
magnetization orientation and $S_0$ is the total spin. Specifically,
\begin{equation} 
\label{chitilde}
\tilde{\chi}^{QP}_{\alpha,\beta}(\omega) = \sum_{ij} \; \frac{f_j-f_i}{\omega_{ij} - \omega - i \eta} \; 
\langle j|s^{\alpha} \Delta_0(\vec{r}) |i\rangle \, \langle i|s^{\beta} \Delta_0(\vec{r})|j\rangle.
\end{equation} 
where $\alpha$ and $\beta$ label the $x$ and $y$ transverse spin directions and 
the easy direction for the magnetization is assummed to be the $\hat{z}$ direction. 
In Eq.(~\ref{chitilde}) $|i\rangle$, $f_{i}$ and $\omega_{ij}$ are Kohn-Sham eigenspinors, Fermi factors, and 
eigenenergy differences respectively, $s_{\alpha}$ is a spin operator, and 
$\Delta_0(\vec{r})$ is the difference between the majority spin and minority spin exchange-correlation 
potential.  In the $s-d$ and $p-d$ models $\Delta_0(\vec{r})$ is replaced by a phenomenological 
constant, which we denote by $\Delta_0$ below.  
With $\Delta_0(\vec{r})$ replaced by a constant $\tilde{\chi}^{QP}_{x,x}$ reduces to a 
standard spin-response function for non-interacting quasiparticles in a 
possibly spin-dependent random static external potential.  The evaluation of this quantity, and 
in particular the low-frequency limit in which we are interested, is non-trivial only
because disorder plays an essential role. 

\subsection{Disorder Perturbation Theory} 

We start by writing the  
transverse spin response function of a disordered metallic ferromagnet in the Matsubara formalism,
\begin{equation}
\tilde{\chi}_{xx}^{QP}(i\omega)=-V\frac{\Delta_0^2}{\beta} \; \sum_{\omega_{n}}P(i\omega_{n},i\omega_{n}+i\omega)
\end{equation}	
where the minus sign originates from fermionic statistics, $V$ is the volume of the system and 
\begin{widetext}
\begin{equation}
\label{eq:kernel}
P(i\omega_{n},i\omega_{n}+i\omega) \equiv  \int\frac{d^{D}k}{(2\pi)^D}\Lambda_{\alpha,\beta}(i\omega_{n},i\omega_{n}+i\omega;\textbf{k})G_{\beta}(i\omega_{n}+i\omega,\textbf{k}) s^{x}_{\beta,\alpha}(\textbf{k}) G_{\alpha}(i\omega_{n},\textbf{k}). 
\end{equation}
In Eq. (~\ref{eq:kernel}) $|\alpha\textbf{k}\rangle $ is a band eigenstate at momentum \textbf{k}, $D$ is the dimensionality of the system, $ s^{x}_{\alpha,\beta}(\textbf{k})= \langle\alpha\textbf{k}|s^{x}|\beta\textbf{k}\rangle $ is the spin-flip matrix element, 
$\Lambda_{\alpha,\beta}(\textbf{k})$ is its vertex-corrected counterpart (see below), and 
\begin{equation}
G_{\alpha}(i\omega_{n},\textbf{k})=\left[i\omega_{n}+E_{F}-E_{\textbf{k},\alpha}+i\frac{1}{2\tau_{\textbf{k},\alpha}}\mbox{sign}(\omega_{n})\right]^{-1} .
\end{equation}
We have included disorder within the Born approximation by incorporating a finite lifetime $\tau$ for the quasiparticles and 
by allowing for vertex corrections at one of the spin vertices.   
\begin{figure}[h]
\begin{center}
\scalebox{0.4}{\includegraphics{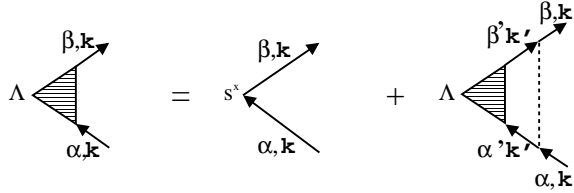}}
\caption{Dyson equation for the renormalized vertex of the transverse spin-spin response function. The dotted line denotes impurity scattering.}
\label{fig:vertex}
\end{center}
\end{figure}

The vertex function in Eq.(~\ref{eq:kernel}) obeys the Dyson equation (Fig. (~\ref{fig:vertex})):
\begin{eqnarray}
\label{eq:dyson}
&&\Lambda_{\alpha,\beta}(i\omega_{n},i\omega_{n}+i\omega;\textbf{k})= s^{x}_{\alpha,\beta}(\textbf{k}) +\nonumber\\
+&&\int\frac{d^{D}k^{\prime}}{(2\pi)^{D}}u^{a}(\textbf{k}-\textbf{k}^{\prime}) s^{a}_{\alpha,\alpha^{\prime}}(\textbf{k},\textbf{k}^{\prime})G_{\alpha^{\prime}}(i\omega_{n},\textbf{k}^{\prime}) \Lambda_{\alpha^{\prime},\beta^{\prime}}(i\omega_{n},i\omega_{n}+i\omega;\textbf{k}^{\prime})
G_{\beta^{\prime}}(i\omega_{n}+i\omega,\textbf{k}^{\prime}) s^{a}_{\beta^{\prime},\beta}(\textbf{k}^{\prime},\textbf{k}),
\end{eqnarray}
\end{widetext}
where $u^{a}(\textbf{q})\equiv n_{a} \overline{V_{a}^{2}}(\textbf{q})(a=0,x,y,z)$, $n_{a}$ is the density of scatterers, $V_{a}(\textbf{q})$ is the scattering potential (dimensions: $\mbox{(energy)}\times\mbox{(volume)}$) and the overline stands for disorder averaging\cite{tatara, disorder}.  Ward's identity
requires that  $u^{a}(\textbf{q})$ and $\tau_{\textbf{k},\alpha}$ be related via the Fermi's golden rule:
\begin{equation}
\label{eq:golden}
\frac{1}{\tau_{\alpha\textbf{k}}}=2\pi\int_{\textbf{k}^{\prime}}u^{a}(\textbf{k}-\textbf{k}^{\prime})\sum_{\alpha^{\prime}}s^{a}_{\alpha,\alpha^{\prime}} s^{a}_{\alpha^{\prime},\alpha} \delta(E_{\textbf{k}\alpha}-E_{\textbf{k}^{\prime}\alpha^{\prime}}),
\end{equation}
where $\int_{\textbf{k}}\equiv \int d^{D} k/(2\pi)^{D}$.
In this paper we restrict ourselves to spin-independent ($a=0$) disorder
and spin-dependent disorder oriented along the equilibrium-exchange-field direction($a=z$)\cite{caveat1}. 
Performing the conventional\cite{mahan} integration around the branch cuts of $P$, we obtain
\begin{widetext}
\begin{equation}
\label{eq:chi_iw}
\tilde{\chi}_{xx}^{QP}(i\omega)= V\Delta_0^2\int_{-\infty}^{\infty}\frac{d\epsilon}{2\pi i}f(\epsilon)\left[P(\epsilon+i\delta,\epsilon+i\omega)-P(\epsilon-i\delta,\epsilon+i\omega)+P(\epsilon-i\omega,\epsilon+i\delta)-P(\epsilon-i\omega,\epsilon-i\delta)\right]
\end{equation}
\end{widetext}
where $f(\epsilon)$ is the Fermi function. Next, we perform an analytical continuation $i\omega \to \omega + i \eta$ 
and take the imaginary part of the resulting retarded response function.  Assuming low temperatures, this yields 
\begin{eqnarray}
\label{eq:alpha_P}
\alpha_{G}&=&\frac{\Delta_{0}^{2}}{2\pi s_{0}}\left\{ \text{Re}\left[P(-i\delta,i\delta)\right]-\text{Re}\left[P(i\delta,+i\delta)\right]\right\} \nonumber\\
&=& \frac{\Delta_{0}^{2}}{2\pi s_{0}}\text{Re}(P^{A,R}-P^{R,R})
\end{eqnarray}
where $s_{0}=S_{0}/V$, 
\begin{equation}
P^{R (A),R} = \int_{\textbf{k}}\Lambda_{\alpha,\beta}^{R (A),R}(\textbf{k})G_{\beta}^{R}(0,\textbf{k}) s^{x}_{\beta,\alpha}(\textbf{k}) G_{\alpha}^{R (A)}(0,\textbf{k}) 
\end{equation}
and $G^{R (A)}(0,\textbf{k})$ is the retarded (advanced)  Green's function at the Fermi energy. 
The principal difficulty of Eq.(~\ref{eq:alpha_P}) resides in solving the Dyson equation for the vertex function. 
We first discuss our method of solution in general terms before turning in Sections III and IV to its application
to the $s-d$ and $p-d$ models. 

\subsection{Evaluation of Impurity Vertex Corrections for Multi-Band Models }
Eq.(~\ref{eq:dyson}) encodes disorder-induced diffusive correlations between itinerant carriers, and is an integral equation of considerable complexity. Fortunately, it is possible to transform it into a relatively simple algebraic equation, provided that the impurity potentials are short-ranged in real space.

Referring back at Eq.(~\ref{eq:dyson}) it is clear that the solution of the Dyson equation would be trivial if the vertex function 
was independent of momentum. That is certainly not the case in general, because the matrix elements of the spin operators may be momentum dependent.  Yet, for short-range scatterers the entire momentum dependence of the vertex matrix elements comes from the eigenstates alone:
\begin{equation}
\label{eq:s_decomposition}
s_{\alpha,\alpha^{\prime}}^{a}(\textbf{k},\textbf{k}^{\prime})=\sum_{m,m^{\prime}}\langle\alpha\textbf{k}|m\rangle\langle m^{\prime}|\alpha^{\prime}\textbf{k}^{\prime}\rangle s_{m,m^{\prime}}^{a}
\end{equation}
This property motivates our solution strategy which characterizes the momentum dependence of the vertex function by expanding it in terms of 
the eigenstates of $s^{z}$ ($s^{x}$ or $s^{y}$ bases would work equally well):
\begin{eqnarray}
\label{eq:decomposition}
\Lambda_{\alpha,\beta}(\textbf{k}) &=& \langle \alpha\textbf{k}|\Lambda|\beta\textbf{k}\rangle\nonumber\\
	&=&\sum_{m,m^{\prime}}\langle\alpha\textbf{k}|m\rangle \Lambda_{m,m^{\prime}}\langle m^{\prime}|\beta\textbf{k}\rangle
\end{eqnarray}
where $|m\rangle$ is an eigenstate of $s^{z}$, with eigenvalue $m$.  Plugging Eqs.(~\ref{eq:s_decomposition}) and (~\ref{eq:decomposition}) into Eq.(~\ref{eq:dyson}) demonstrates that, as expected, $\Lambda_{m,m^{\prime}}$ is \emph{independent} of momentum.
After cancelling common factors from both sides of the resulting expression and using $\partial_{\textbf{q}} u^{a}(\textbf{q})=0$ ($a=0,z)$ we arrive at
\begin{equation}
\label{eq:matrix equation}
\Lambda_{m,m^{\prime}}^{R(A),R}=s_{m,m^{\prime}}^{x}+\sum_{l,l^{\prime}} U_{m, m^{\prime}: l, l^{\prime}}^{R(A),R} \Lambda_{l,l^{\prime}}^{R(A),R}
\end{equation}
where
\begin{widetext}
\begin{equation}
\label{eq:U-matrix}
U_{m,m^{\prime}:l,l^{\prime}}^{R(A),R}\equiv \left(u^{0}+ u^{z} m m^{\prime}\right)\int_{\textbf{k}} \langle m|\alpha\textbf{k}\rangle G_{\alpha}^{R(A)}(0,\textbf{k})\langle\alpha\textbf{k}|l\rangle \langle l^{\prime}|\beta\textbf{k}\rangle G_{\beta}^{R}(0,\textbf{k})\langle\beta\textbf{k}|m^{\prime}  \rangle
\end{equation}
\end{widetext}
Eqs. (~\ref{eq:decomposition}),(~\ref{eq:matrix equation}) and (~\ref{eq:U-matrix}) provide a solution for the vertex function that is significantly easier to analyse than the original Dyson equation. 

\section{Gilbert Damping for a Magnetic 2DEG}

The first model we consider is a two-dimensional electron gas (2DEG) model with ferromagnetism and Rashba 
spin-orbit interactions.  We refer to this as the magnetic 2DEG (M2DEG) model.
This toy model is almost never even approximately realistic\cite{furdyna}, but a theoretical study of 
its properties will prove useful in a number of ways.
First, it is conducive to a fully analytical evaluation of the Gilbert damping, which will allow us to precisely understand the role of different actors. Second, it enables us to explain in simple terms why higher order vertex corrections are significant when there is spin-orbit interaction in the band structure. Third, the Gilbert damping of a M2DEG has qualitative features similar to those of (Ga,Mn)As.
 
The band Hamiltonian of the M2DEG model is
\begin{equation}
H=\frac{k^{2}}{2 m}+\textbf{b}_{\textbf{k}}\cdot\mathbf{\sigma}
\end{equation}
where $\textbf{b}_{\textbf{k}}=(-\lambda k_{y}, \lambda k_{x}, \Delta_0)$, $\Delta_0$ is the difference between 
majority and minority spin exchange-correlation potentials, $\lambda$ is the strength of the Rashba SO coupling and $\vec{\sigma}=2\vec{s}$ is a vector of Pauli matrices. 
The corresponding eigenvalues and eigenstates are
\begin{eqnarray}
&& E_{\pm,\textbf{k}}=\frac{k^2}{2m}\pm \sqrt{\Delta_0^{2}+\lambda^{2}k^{2}}\\
\label{eq:eigenstate}
&&|\alpha\textbf{k}\rangle = e^{-i s^{z}\phi} e^{-i s^{y}\theta}|\alpha\rangle
\end{eqnarray}
where $\phi=-\mbox{tan}^{-1}(k_{x}/k_{y})$ and $ \theta=\mbox{cos}^{-1}(\Delta_0/\sqrt{\Delta_0^{2}+\lambda^{2}k^{2}})$ are the spinor angles and $\alpha=\pm$ is the band index.
It follows that 
\begin{eqnarray}
\label{eq:matrix_element}
\langle m|\alpha,\textbf{k}\rangle &=& \langle m|e^{-i s^{z}\phi} e^{-i s^{y}\theta}|\alpha\rangle\nonumber\\
&=& e^{-i m\phi} d_{m,\alpha}(\theta)
\end{eqnarray}
where $d_{m,\alpha}=\langle m|e^{-i s^{y}\theta}|\alpha\rangle$ is a Wigner function for J=1/2 angular momentum\cite{sakurai}.
With these simple spinors, the azimuthal integral in Eq.(~\ref{eq:U-matrix}) can be performed analytically to obtain
\begin{widetext}
\begin{equation}
\label{eq:U_M2DEG_2}
U_{m, m^{\prime}: l, l^{\prime}}^{R(A),R} = \delta_{m-m^{\prime},l-l^{\prime}} (u^{0}+u^{z} m m^{\prime})\sum_{\alpha,\beta}\int\frac{dk k}{2\pi} d_{m\alpha} G_{\alpha}^{R(A)}(k) d_{l\alpha}(\theta) d_{m^{\prime}\beta}(\theta) G_{\beta}^{R}(k)  d_{l^{\prime}\beta}(\theta),  
\end{equation}
\end{widetext}
where the Kronecker delta reflects the conservation of the angular momentum along $z$, owing to the azimuthal symmetry of the problem. 
In Eq.(~\ref{eq:U_M2DEG_2}) 
\begin{equation}
\label{eq:wigner}
d_{m,m^{\prime}}=\left(\begin{array}{cc} \cos(\theta/2) & -\sin(\theta/2) \\ \sin(\theta/2) & \cos(\theta/2)\end{array}\right),
\end{equation}
and the retarded and advanced Green's functions are 
\begin{eqnarray}
G_{+}^{R(A)}&=& \frac{1}{-\xi_{k}-b_{k}+ (-) i\gamma_{+}}\nonumber\\
G_{-}^{R(A)}&=& \frac{1}{-\xi_{k}+b_{k}+ (-) i\gamma_{-}},
\end{eqnarray}
where $\xi_{k}=\frac{k^{2}-k_{F}^{2}}{2 m}$, $b_{k}=\sqrt{\Delta_0^{2}+\lambda^{2}k^{2}}$,
and $\gamma_{\pm}$ is (half) the golden-rule scattering rate of the band quasiparticles.
In addition, Eq. (~\ref{eq:matrix equation}) is readily inverted to yield  
\begin{eqnarray}
\label{eq:vertex_M2DEG_2}
\Lambda_{+,+}^{R(A),R}&=&\Lambda_{-,-}^{R(A),R}=0\nonumber\\
\Lambda_{+,-}^{R(A),R}&=&\frac{1}{2}\frac{1}{1-U_{+,-:+,-}^{R(A),R}}\nonumber\\
\Lambda_{-,+}^{R(A),R}&=& \frac{1}{2}\frac{1}{1-U_{-,+:-,+}^{R(A),R}}
\end{eqnarray}

In order to make further progress analytically we assume that $(\Delta_0,\lambda k_{F}, \gamma)<<E_{F}=k_{F}^{2}/2m$.
It then follows that  $\gamma_{+}\simeq \gamma_{-}\equiv\gamma$
and that $\gamma=\pi N_{2D} u^{0}+\pi N_{2D} \frac{u^{z}}{4}\equiv \gamma_{0}+\gamma_{z}$. 
Eqs. (~\ref{eq:U_M2DEG_2}) and (~\ref{eq:wigner}) combine to give  
\begin{widetext}
\begin{eqnarray}
\label{eq:U_M2DEG_4}
U_{-,+:-,+}^{R,R}&=& U_{+,-:+,-}^{R,R}=0 \mbox{  }  
\nonumber\\
U_{-,+:-,+}^{A,R}&=&(\gamma_{0}-\gamma_{z}) \left[\frac{i}{-b+i\gamma}\cos^{4}\left(\frac{\theta}{2}\right)+\frac{i}{b+i\gamma}\sin^{4}\left(\frac{\theta}{2}\right)+\frac{2}{\gamma} \cos^{2}\left(\frac{\theta}{2}\right) \sin^{2}\left(\frac{\theta}{2}\right)\right]\nonumber\\
U_{+,-:+,-}^{A,R}&=&(U_{-,+:-,+}^{A,R})^{\star}
\end{eqnarray}
\end{widetext}
where $b\simeq\sqrt{\lambda^{2}k_{F}^{2}+\Delta_0^{2}}$ and $\cos\theta\simeq \Delta_0/b$. The first and second terms in square brackets in Eq.(~\ref{eq:U_M2DEG_4}) emerge from inter-band transitions ($\alpha\neq\beta$ in Eq. (~\ref{eq:U_M2DEG_2})), while the last term stems from intra-band transitions ($\alpha=\beta$). Amusingly, $U$ vanishes when the spin-dependent scattering rate equals the Coulomb scattering rate ($\gamma_{z}=\gamma_{0}$); in this particular instance vertex corrections are completely absent.  On the other hand,
when $\gamma_{z}=0$ and $b<<\gamma$ we have $U_{-,+:-,+}^{A,R}\simeq U_{+,-:+,-}^{A,R}\simeq 1$, implying that 
vertex corrections strongly enhance Gilbert damping (recall Eq. (~\ref{eq:vertex_M2DEG_2})).  We will discuss the role of vertex corrections more fully below.

\begin{figure}[h]
\begin{center}
\scalebox{0.4}{\includegraphics{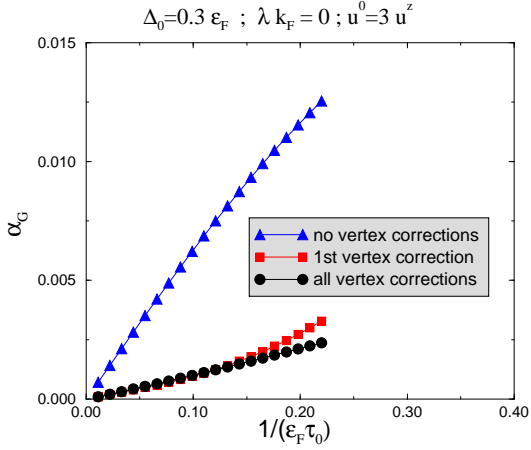}}
\caption{\textbf{M2DEG}: Gilbert damping in the absence of spin-orbit coupling.
When the intrinsic spin-orbit interaction is small, the 1st vertex correction is sufficient for the evaluation of Gilbert damping, provided that
the ferromagnet's exchange splitting is large compared to the lifetime-broadening of the quasiparticle energies. For more disordered ferromagnets ($E_{F}\tau_{0}<5$ in this figure) higher order vertex corrections begin to matter. In either case vertex corrections 
are significant. In this figure $1/\tau_{0}$ stands for the scattering rate off spin-independent impurities,
defined as a two-band average at the Fermi energy, and the spin-dependent and spin-independent impurity strengths 
are chosen to satisfy $u^{0}=3 u^{z}$.}
\label{fig:M2DEG1}
\end{center}
\end{figure}

\begin{figure}[h]
\begin{center}
\scalebox{0.4}{\includegraphics{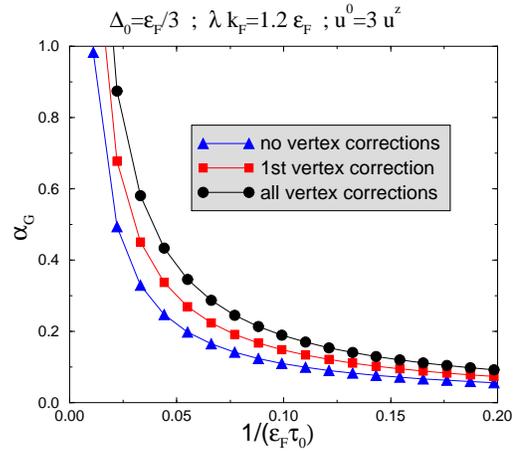}}
\caption{\textbf{M2DEG}: Gilbert damping for strong SO interactions ($\lambda k_{F}=1.2 E_F\simeq 4 \Delta_0$).  In this case 
higher order vertex corrections matter (up to 20 \%) even at low disorder.
This suggests that higher order vertex corrections will be important in real
ferromagnetic semiconductors because their intrinsic SO interactions are 
generally stronger than their exchange splittings.}
\label{fig:M2DEG2}
\end{center}
\end{figure}
 \begin{figure}[h]
\begin{center}
\scalebox{0.4}{\includegraphics{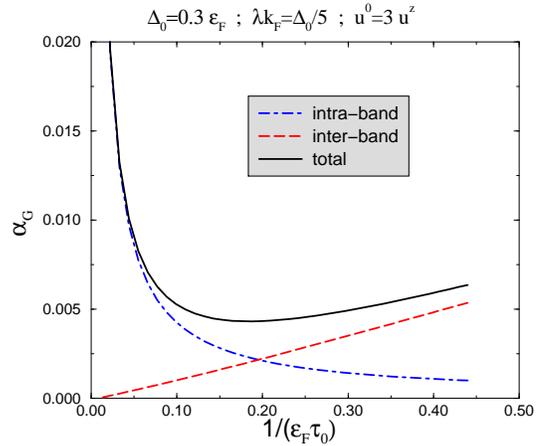}}
\caption{\textbf{M2DEG}: Gilbert damping for moderate SO interactions ($\lambda k_{F}=0.2 \Delta_0$).   In this case 
there is a crossover between the intra-band dominated and the inter-band dominated regimes, which gives rise to a non-monotonic 
dependence of Gilbert damping on disorder strength. The stronger the intrinsic SO relative to the exchange field, the higher the value of disorder at which the crossover occurs.  This is why the damping is monotonically increasing with disorder in Fig. (~\ref{fig:M2DEG1}) and monotonically decreasing in Fig. (~\ref{fig:M2DEG2}).}
\label{fig:M2DEG3}
\end{center}
\end{figure}

After evaluating $\Lambda (\textbf{k})$ from Eqs. (~\ref{eq:decomposition}),(~\ref{eq:vertex_M2DEG_2})and (~\ref{eq:U_M2DEG_4}), the last step is to compute
\begin{equation}
\label{eq:kernel_M2DEG}
P^{R(A),R}=\int_{\textbf{k}}\Lambda_{\alpha,\beta}^{R(A), R}(\textbf{k}) s_{\beta,\alpha}^{x}(\textbf{k}) G_{\alpha}^{R(A)}(\textbf{k}) G_{\beta}^{R}(\textbf{k}).\\
\end{equation}
Since we are assuming that the Fermi energy is the largest energy scale, the integrand in Eq. (~\ref{eq:kernel_M2DEG}) is sharply peaked at the Fermi surface, leading to $P^{R,R}\simeq 0$.
In the case of spin-independent scatterers ($\gamma_{z}=0\rightarrow \gamma=\gamma_{0}$), tedious but straightforward algebra takes us to
\begin{equation}
\label{eq:ewelina}
\alpha_{G} (u^{z}=0)=\frac{ N_{2D}\Delta_0^2}{4 s_0\gamma_0}\frac{(\lambda^{2}k_{F}^{2})(b^{2}+\Delta_0^{2}+2 \gamma_{0}^{2})}{(b^{2}+\Delta_0^{2})^{2}+4 \Delta_0^{2}\gamma_{0}^{2}}.
\end{equation}
Eq. (29) agrees with results published in the recent literature\cite{ewelina}. We note that $\alpha_{G}(u^{z}=0)$ vanishes in 
the absence of SO interactions, as expected. It is illustrative to expand Eq. (~\ref{eq:ewelina}) in the $b>>\gamma_{0}$ regime:
\begin{equation}
\label{eq:spinindep}
\alpha_{G} (u^{z}=0)\simeq\frac{N_{2D}\Delta_0^2 }{2 s_0}\left[\frac{\lambda^{2} k_{F}^{2}}{2(b^{2}+\Delta_0^{2})}\frac{1}{\gamma_{0}}+ \frac{\lambda^{4} k_{F}^{4}}{(b^{2}+\Delta_0^{2})^{3}}\gamma_{0}\right]
\end{equation}
which displays intra-band $(\sim\gamma_{0}^{-1})$ and inter-band $(\sim\gamma_{0})$ contributions separately.
The intra-band damping is due to the dependence of band eigenenergies on magnetization orientation, the 
{\em breathing Fermi surface} effect\cite{oldkambersky} which produces more damping when the band-quasiparticles
scatter infrequently because the population distribution moves further from equilibrium.  The intra-band contribution
to damping therefore tends to scale with the conductivity.   
For stronger disorder, the inter-band term in which scattering relaxes spin-orientations
takes over and $\alpha_{G}$ is proportional to the resistivity. 
Insofar as phonon-scattering can be treated as elastic, the Gilbert damping will often show a non-monotonic temperature dependence
with the intra-band mechanism dominating at low-temperatures when the conductivity is large and the 
inter-band mechanism dominating at high-temperatures when the resistivity is large.

For completeness, we also present analytic results for the case $\gamma=\gamma_{z}$ in the $b>>\gamma_{z}$ regime:
\begin{equation}
\label{eq:spindep}
\alpha_{G}(u^{0}=0)\simeq\frac{N_{2D}\Delta_0^2}{2 s_0}\left[\frac{1}{\gamma_{z}}\frac{\lambda^{2} k_{F}^{2}}{6 b^{2}-2 \Delta_0^{2}}+\gamma_{z}\frac{3 b^{4}+6 b^{2} \Delta_0^{2}-\Delta_0^{4}}{(3 b^{2}-\Delta_0^{2})^{3}}\right]
\end{equation}
This expression illustrates that spin-orbit (SO) interactions in the band structure are 
a necessary condition for the intra-band transition contribution to $\alpha_{G}$.  
The interband contribution survives in absence of SO as long as the disorder potential 
is spin-dependent.
Interband scattering is possible for spin-dependent disorder because 
majority and minority spin states on the Fermi surface are not orthogonal
when their potentials are not identical.
Note incidentally the contrast between Eq.(~\ref{eq:spinindep}) and Eq. (~\ref{eq:spindep}): in the former the inter-band coefficient is most suppressed at weak intrinsic SO interaction  while in the latter it is the intra-band coefficient which gets weakest for small $\lambda k_{F}$.

More general cases relaxing the $(\Delta_0,\lambda k_{F},\gamma)<<E_F$ assumption must be studied numerically; the results are collected in Figs. (~\ref{fig:M2DEG1}), (~\ref{fig:M2DEG2}) and (~\ref{fig:M2DEG3}). 
Fig (~\ref{fig:M2DEG1}) highlights the inadequacy of completely neglecting vertex corrections in the limit of weak spin-orbit interaction; the inclusion of the the leading order vertex correction largely solves the problem. However, 
Fig. (~\ref{fig:M2DEG1}) and (~\ref{fig:M2DEG2}) together indicate that higher order vertex corrections are noticeable 
when disorder or spin-orbit coupling is strong. 
In the light of the preceding discussion the monotonic decay in Fig.(~\ref{fig:M2DEG2}) may appear surprising because the inter-band contribution presumably increases with $\gamma$. Yet, this argument is strictly correct only for weakly spin-orbit coupled systems, where the crossover betwen inter-band and intra-band dominated regimes occurs at low disorder.  For strongly spin-orbit coupled systems the crossover may take place at a scattering rate that is (i) beyond experimental relevance and/or (ii) larger than the band-splitting, in which case the inter-band contribution behaves much like its intra-band partner, i.e. $O(1/\gamma)$. Non-monotonic behavior is restored when the spin-orbit splitting is weaker, as shown in Fig. (~\ref{fig:M2DEG3}).

Finally, our analysis opens an opportunity to quantify the importance of higher order impurity vertex-corrections. 
Kohno, Shibata and Tatara \cite{tatara} claim that the bare vertex along with the \emph{first} vertex correction fully captures the Gilbert damping of a ferromagnet, provided that $\Delta_0\tau>>1$. 
To first order in $U$ the vertex function is  
\begin{equation}
\Lambda_{m, m^{\prime}}^{R(A), R}=s_{m, m^{\prime}}^{x}+\sum_{l l^{\prime}} U_{m, m^{\prime}: l, l^{\prime}}^{R(A),R} s^{x}_{l, l^{\prime}}
\end{equation}
Taking $\gamma=\gamma_{z}$ for simplicity, we indeed get
\begin{eqnarray}
\label{eq:A}
&&\lim_{\lambda\rightarrow 0}\alpha_{G}\simeq A\gamma +O(\gamma^{2})\nonumber\\
&&\frac{A(1)}{A(\infty)}=1\mbox{      }
\end{eqnarray}
where $A(1)$ contains the first vertex correction only, and $A(\infty)$ includes all vertex corrections. However, the state of affairs  changes after turning on the intrinsic SO interaction, whereupon Eq. (~\ref{eq:A}) transforms into
\begin{eqnarray}
&&\alpha_{G}(\lambda\neq 0)\simeq B\gamma+C\frac{1}{\gamma} \nonumber\\
&&\frac{B(1)}{B(\infty)}=
\frac{\Delta_0^{2}(3 b^{2}-\Delta_0^{2})^{3}(3 b^{2}+\Delta_0^{2})}{4 b^{6}(3 b^{4}+6 b^{2}\Delta_0^{2}-\Delta_0^{4})}\nonumber\\
&&\frac{C(1)}{C(\infty)}=
\frac{(b^{2}+\Delta_0^{2})(3 b^{2}-\Delta_0^{2})}{4 b^{4}}
\end{eqnarray}
When $\Delta_0<<\lambda k_{F}$, \emph{both} intra-band and inter-band ratios show a significant deviation from unity \cite{B and C}, to which they converge as $\lambda\rightarrow 0$. In order to understand this behavior, let us look back at Eq. (~\ref{eq:vertex_M2DEG_2}). 
There, we can formally expand the vertex function as $\Lambda=\frac{1}{2}\sum_{n=0}^{\infty} U^{n}$, where the $n$-th order term stems from the $n$-th vertex correction. From Eq. (~\ref{eq:U_M2DEG_4}) we find that when $\lambda=0$, $U^{n}\sim O(\gamma^{n})$ and thus $n\geq 2$ vertex corrections will not matter for the Gilbert damping, which is $O(\gamma)$\cite{greens} when $E_{F}>>\gamma$. In contrast, when $\lambda\neq 0$ the intra-band term in Eq. (~\ref{eq:U_M2DEG_4}) is no longer zero, and consequently \emph{all} powers of $U$ contain $O(\gamma^{0})$ and $O(\gamma^{1})$ terms. In other words, all vertices contribute to $O(1/\gamma)$ and $O(\gamma)$ in the Gilbert damping, especially if $\lambda k_{F}/\Delta_0$ is not small. This conclusion should prove valid beyond the realm of the M2DEG because it relies only on the mantra ``intra-band$\sim O(1/\gamma)$; inter-band $\sim O(\gamma)$''. Our expectation that higher order vertex corrections be important in (Ga,Mn)As will be confirmed numerically in the next section.  

\section{Gilbert Damping for $\text{(Ga,Mn)As}$}

(Ga,Mn)As and other (III,Mn)V ferromagnets are like transition metals in that their 
magnetism is carried mainly by d-orbitals, but unlike transition metals in that
neither majority nor minority spin d-orbitals are present at the Fermi energy.
The orbitals at the Fermi energy 
are very similar to the states near the top of the valence band states of the host (III,V) semiconductor,
although they are of course weakly hybridized with the minority and majority 
spin d-orbitals.  For this reason the electronic structure of (III,Mn)V 
ferromagnets is extremely simple and can be described reasonably accurately
with the phenomenological model which we employ in this section.  Because 
the top of the valence band in (III,V) semiconductors is split by 
spin-orbit interactions, spin-orbit coupling plays a dominant role in the 
bands of these ferromagnets.    
An important consequence of the strong SO interaction in the band structure is that diffusive vertex corrections 
influence $\alpha_{G}$ significantly at \emph{all} orders; this is the central idea of this section.

Using a p-d mean-field theory model\cite{gamnas} for the ferromagnetic ground state and a four-band spherical model\cite{cardona} for the host semiconductor band structure, $\mbox{Ga}_{1-x} \mbox{Mn}_{x} \mbox{As}$ may be described by
\begin{equation}
H=\frac{1}{2 m}\left[\left(\gamma_{1}+\frac{5}{2}\gamma_{2}\right)k^{2}-2\gamma_{3}(\textbf{k}\cdot\textbf{s})^{2}\right]+\Delta_0 s^{z},
\end{equation}
where $\textbf{s}$ is the spin operator projected onto the J=3/2 total angular momentum subspace at the top of the valence band and \{$\gamma_{1}=6.98, \gamma_{2}=\gamma_{3}=2.5$\} are the Luttinger parameters for 
the spherical-band approximation to GaAs. In addition, $\Delta_0=J_{pd} S N_{Mn}$ is the exchange field, $J_{pd}=55 \mbox{meV} \mbox{nm}^{3}$ is the p-d exchange coupling, $S=5/2$ is the spin of the Mn ions, $N_{Mn}=4x/a^{3}$ is the density of Mn ions, and $a=0.565 \mbox{nm}$ is the lattice constant of GaAs.

\begin{figure}[h]
\begin{center}
\scalebox{0.4}{\includegraphics{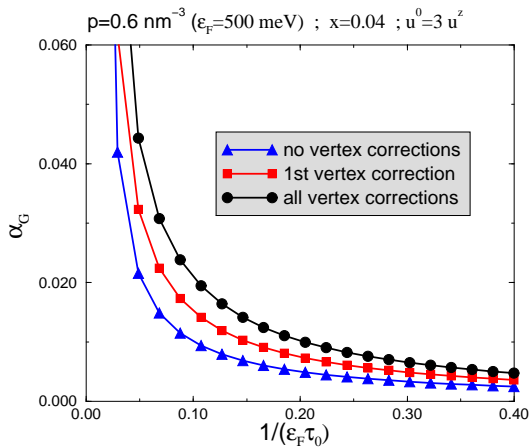}}
\caption{\textbf{GaMnAs}: Higher order vertex corrections make a significant contribution to Gilbert damping, 
due to the prominent spin-orbit interaction in the band structure of GaAs. $x$ is the Mn fraction, and $p$ is the hole concentration that determines the Fermi energy $E_{F}$. 
In this figure, the spin-independent impurity strength $u^{0}$ was taken to be 3 times larger than the magnetic impurity strength $u^{z}$. $1/\tau_{0}$ corresponds to the scattering rate off Coulomb impurities and is evaluated as a four-band average at the Fermi energy.}
\label{fig:GaMnAs1}
\end{center}
\end{figure}

\begin{figure}[h]
\begin{center}
\scalebox{0.4}{\includegraphics{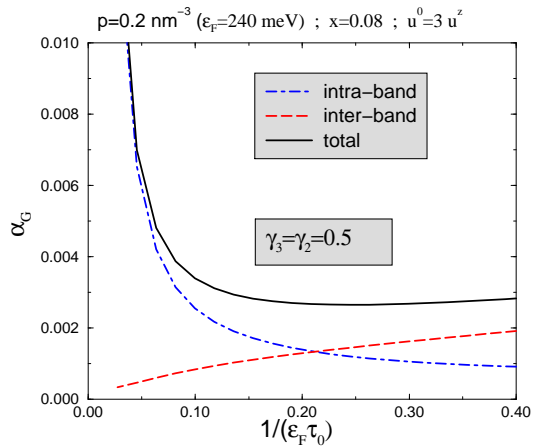}}
\caption{\textbf{GaMnAs}: When the spin-orbit splitting is reduced (in this case by reducing the hole density to $0.2 nm^{-3}$ and artificially taking $\gamma_{3}=0.5$), the crossover between inter- and intra-band dominated regimes produces a non-monotonic shape of the Gilbert damping, 
much like in Fig. (~\ref{fig:M2DEG3}). When either $\gamma_{2}$ or $p$ is made larger or $x$ is reduced, 
we recover the monotonic decay of Fig.(~\ref{fig:GaMnAs1}).}
\label{fig:GaMnAs2}
\end{center}
\end{figure}
 
The $\Delta_0=0$ eigenstates of this model are
\begin{equation}
\label{eq:eigenstates_GaAs}
|\tilde{\alpha},\textbf{k}\rangle=e^{-i s^{z}\phi} e^{-i s^{y}\theta}|\tilde{\alpha}\rangle
\end{equation}
where $|\tilde{\alpha}\rangle$ is an eigenstate of $s^{z}$ with eigenvalue $\tilde{\alpha}$. Unfortunately, the analytical form of the $\Delta_0\neq 0$ eigenstates is unknown. Nevertheless, since the exchange field preserves the azimuthal symmetry of the problem, the $\phi$-dependence of the full eigenstates $|\alpha\textbf{k}\rangle$ will be identical to that of Eq. (~\ref{eq:eigenstates_GaAs}). This observation leads to $U_{m,m^{\prime}:l,l^{\prime}}\propto\delta_{m-m^{\prime},l-l^{\prime}}$, which simplifies Eq. (~\ref{eq:U-matrix}).
$\alpha_{G}$ can be calculated numerically following the steps detailed in the previous sections; 
the results are summarized in Figs. (~\ref{fig:GaMnAs1}) and (~\ref{fig:GaMnAs2}). 
Note that vertex corrections moderately increase the damping rate, as in the case of a M2DEG model with strong
spin-orbit interactions.
Fig. (~\ref{fig:GaMnAs1}) underlines both the importance of higher order vertex corrections in (Ga,Mn)As and the monotonic decay of the damping as a function of scattering rate. The latter signals the supremacy of the intra-band contribution to damping, 
accentuated at larger hole concentrations. 
Had the intrinsic spin-orbit interaction been substantially weaker\cite{weak SO}, $\alpha_{G}$ would have traced a non-monotonic curve as shown in Fig. (~\ref{fig:GaMnAs2}).  The degree to which the intraband {\em breathing Fermi surface} model effect dominates 
depends on the details of the band-structure and can be influenced by corrections to the spherical model which we 
have adopted here to simplify the vertex-correction calculation.  The close 
correspondence between Figs. (~\ref{fig:GaMnAs1})-(~\ref{fig:GaMnAs2}) and Figs. (~\ref{fig:M2DEG2})-(~\ref{fig:M2DEG3}) reveals the success of the M2DEG as a versatile gateway for realistic models and justifies the extensive attention devoted to it in this paper and elsewhere.

\section{Assessment of the torque-correlation formula}
Thus far we have evaluated the Gilbert damping for a M2DEG model and a (Ga,Mn)As model
using the (bare) spin-flip vertex $\langle\alpha,\textbf{k}|s^{x}|\beta,\textbf{k}\rangle$ and its renormalized counterpart $\langle\alpha,\textbf{k}|\Lambda|\beta,\textbf{k}\rangle$.
The vertex corrected results are expected to be exact for $1/\tau$ small compared to the Fermi energy. 
For practical reasons, state-of-the-art band-structure calculations\cite{realistic} forgo impurity vertex corrections altogether and instead employ the \emph{torque-correlation} matrix element, which we shall denote as $\langle\alpha,\textbf{k}|K|\beta,\textbf{k}\rangle$ (see below for an explicit expression). In this section we compare damping rates calculated using $s^{x}_{\alpha,\beta}$ vertices with those 
calculated using $K_{\alpha,\beta}$ vertices.  We also compare both results with the exact damping rates 
obtained by using $\Lambda_{\alpha,\beta}$. The ensuing discussion overlaps with and extends our recent preprint\cite{alphaI}.

We shall begin by introducing the following identity\cite{oldkambersky}:
\begin{eqnarray}
\label{eq:n_vs_k}
\langle \alpha,\textbf{k}|s^{x}|\beta,\textbf{k}\rangle\nonumber &=&i \langle \alpha,\textbf{k}|\left[s^{z},s^{y}\right]|\beta,\textbf{k}\rangle \nonumber\\
&=& \frac{i}{\Delta_0}(E_{\textbf{k},\alpha}-E_{\textbf{k},\beta})\langle \alpha,\textbf{k}|s^{y}|\beta,\textbf{k}\rangle\nonumber\\
&-&\frac{i}{\Delta_0}\langle \alpha,\textbf{k}|\left[H_{so},s^{y}\right]|\beta,\textbf{k}\rangle. 
\end{eqnarray}
In Eq. (~\ref{eq:n_vs_k}) we have decomposed the mean-field quasiparticle Hamiltonian into a sum of spin-independent,
exchange spin-splitting, and other spin-dependent terms: $H=H_{kin}+H_{so}+H_{ex}$, where $H_{kin}$ is the kinetic (spin-independent) part,
$H_{ex}=\Delta_0 s^{z}$ is the exchange spin-splitting term and $H_{so}$ is the piece that contains the intrinsic spin-orbit interaction. 
The last term on the right hand side of Eq. (~\ref{eq:n_vs_k}) is the torque-correlation matrix element used in band structure computations:
\begin{equation}
\langle \alpha,\textbf{k}|K|\beta,\textbf{k}\rangle\equiv-\frac{i}{\Delta_0}\langle \alpha,\textbf{k}|\left[H_{so},s^{y}\right]|\beta,\textbf{k}\rangle. 
\end{equation} 
Eq. (~\ref{eq:n_vs_k}) allows us to make a few general remarks on the relation between the spin-flip and torque-correlation matrix elements. 
For intra-band matrix elements, one immediately finds that $s^{x}_{\alpha,\alpha}=K_{\alpha,\alpha}$ and hence the two approaches agree.  
For inter-band matrix elements the agreement between $s^{x}_{\alpha,\beta}$ and $K_{\alpha,\beta}$ should be 
nearly identical when the first term in the final form of Eq.(~\ref{eq:n_vs_k}) is small,
{\em i.e.} when\cite{caveat2} $(E_{\textbf{k},\alpha}-E_{\textbf{k},\beta})<<\Delta_0$. Since this requirement cannot be satisfied in the M2DEG, we expect that the inter-band contributions from $K$ and $s^{x}$ will always differ significantly in this model. 
More typical models, like the four-band model for (Ga,Mn)As, have band crossings at a discrete set of k-points, 
in the neighborhood of which $K_{\alpha,\beta}\simeq s^{x}_{\alpha,\beta}$. 
The relative weight of these crossing points in the overall Gilbert damping depends on a variety of factors. 
First, in order to make an impact they must be located within a shell of thickness $1/\tau$ around the Fermi surface. 
Second, the contribution to damping from those special points must outweigh that from the remaining k-points in the shell; this might be the case for instance in materials with weak spin-orbit interaction and weak disorder, where the contribution from the crossing points would go like $\tau$ (large) while the contribution from points far from the crossings would be $\sim1/\tau$ (small). Only if these two conditions are fulfilled should one expect
good agreement between the inter-band contribution from spin-flip and torque-correlation formulas.
When vertex corrections are included, of course, the same result should be obtained using either form for 
the matrix element, since all matrix elements are between essentially degenerate electronic states
when disorder is treated non-perturbatively\cite{alphaI,ewelina}.
\begin{figure}[h]
\begin{center}
\scalebox{0.4}{\includegraphics{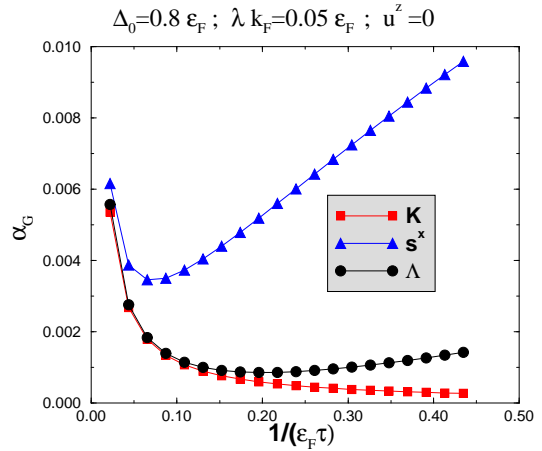}}
\caption{\textbf{M2DEG}: Comparison of Gilbert damping predicted using spin-flip and torque matrix 
element formulas, as well as the exact vertex corrected result.
In this figure the intrinsic spin-orbit interaction is relatively weak ($\lambda k_{F}=0.05 E_F\simeq 0.06 \Delta_0$)
and we have taken $u^{z}=0$. The torque correlation formula does not distinguish 
between spin-dependent and spin-independent disorder.}
\label{fig:M2DEG_comp1}
\end{center}
\end{figure}

\begin{figure}[h]
\begin{center}
\scalebox{0.4}{\includegraphics{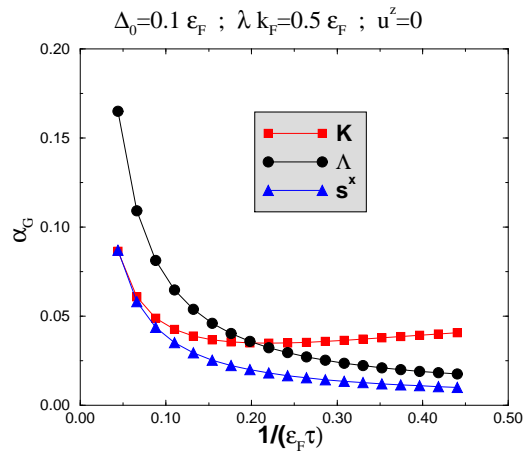}}
\caption{\textbf{M2DEG}: Comparison of Gilbert damping predicted using spin-flip and torque matrix 
element formulas, as well as the exact vertex corrected result.
In this figure the intrinsic spin-orbit interaction is relatively strong ($\lambda k_{F}=0.5 E_F= 5 \Delta_0$)
and we have taken $u^{z}=0$}
\label{fig:M2DEG_comp2}
\end{center}
\end{figure}

In the remining part of this section we shall focus on a more quantitative comparison between the different formulas. 
For the M2DEG it is straightforward to evaluate $\alpha_G$ analytically using $K$ instead of $s^{x}$ and neglecting vertex corrections; we obtain
\begin{equation}  
\label{eq:alpha_k}
\alpha_{G}^{K}=\frac{N_{2D}\Delta_0}{8 s_0}\left[\frac{\lambda^{2} k_{F}^{2}}{b^{2}}\frac{\Delta_0}{\gamma}+\left(\frac{\lambda^{2}k_{F}^{2}}{\Delta_0 b}\right)^{2}\frac{\gamma \Delta_0}{\gamma^{2}+b^{2}}\right]
\end{equation} 
where we assumed $(\gamma, \lambda k_{F}, \Delta_0)<<\epsilon_{F}$. By comparing Eq. (~\ref{eq:alpha_k}) with the exact expression Eq. (~\ref{eq:ewelina}), we find that the intra-band parts are in excellent agreement when $\Delta_0<<\lambda k_{F}$, i.e. when vertex corrections are relatively unimportant.  
In contrast,  the inter-band parts differ markedly regardless of the vertex corrections. These trends are captured by Figs. (~\ref{fig:M2DEG_comp1}) and (~\ref{fig:M2DEG_comp2}), which compare the Gilbert damping obtained from $s^{x}$, $K$ and $\Lambda$ matrix elements. 
Fig. (~\ref{fig:M2DEG_comp1}) corresponds to the weak spin-orbit limit, where it is found that in disordered ferromagnets $s^{x}$ may grossly overestimate the Gilbert damping because its inter-band contribution does not vanish even as SO tends to zero. 
As explained in Section III, this flaw may be repaired by adding the leading order impurity vertex correction. The torque-correlation formula is free from such problem because $K$ vanishes identically in absence of SO interaction. Thus the main practical advantage of $K$ is that it yields a physically sensible result without having to resort to vertex corrections. Continuing with Fig.(~\ref{fig:M2DEG_comp1}), at 
weak disorder the intra-band contributions dominate and therefore $s^{x}$ and $K$ coincide; even $\Lambda$ agrees, because for intra-band transitions at weak spin-orbit interaction the vertex corrections are unimportant. Fig. (~\ref{fig:M2DEG_comp2}) corresponds to the strong spin-orbit case. In this case,  at low disorder $s^{x}$ and $K$ agree well with each other, but differ from the exact result because higher order vertex corrections alter the intra-band part substantially. For a similar reason, neither $s^{x}$ nor $K$ agree with the exact $\Lambda$ at higher disorder.  
Based on these model calculations, we do not believe that there are any objective grounds to 
prefer either the $K$ torque-correlation or the $s^{x}$ spin-flip formula estimate of $\alpha_{G}$ when 
spin-orbit interactions are strong and $\alpha_{G}$ is dominated by 
inter-band relaxation. 
A precise estimation of $\alpha_{G}$ under these circumstances appears to require that the 
character of disorder, incuding its spin-dependence, be accounted for reliably and that the 
vertex-correction Dyson equation be accurately solved.  Carrying out this program remains a challenge both 
because of technical complications in performing the calculation for general band structures and 
because disorder may not be sufficiently well characterized. 

\begin{figure}[h]
\begin{center}
\scalebox{0.4}{\includegraphics{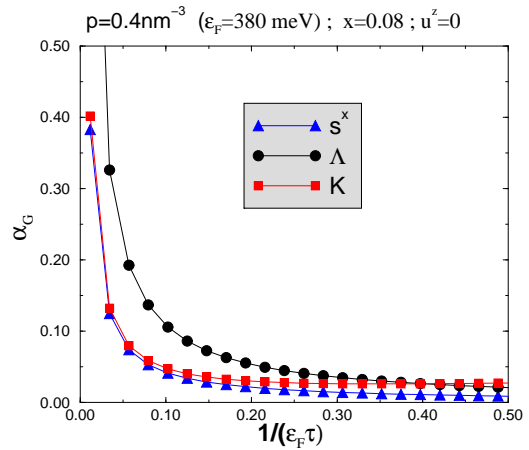}}
\caption{\textbf{GaMnAs}: Comparison of Gilbert damping predicted using spin-flip and torque matrix 
element formulas, as well as the exact vertex corrected result. $p$ is the hole concentration that determines the Fermi energy $E_{F}$ and $x$ is the Mn fraction. Due to the strong intrinsic SO, this figure shows similar features as Fig.(~\ref{fig:M2DEG_comp2}).}
\label{fig:GaMnAs_comp1}
\end{center}
\end{figure}

\begin{figure}[h]
\begin{center}
\scalebox{0.4}{\includegraphics{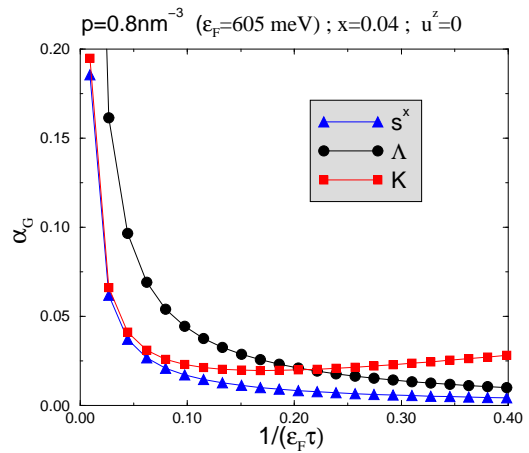}}
\caption{\textbf{GaMnAs}: Comparison of Gilbert damping predicted using spin-flip and torque matrix 
element formulas, as well as the exact vertex corrected result. In relation to Fig. (~\ref{fig:GaMnAs_comp1}) the effective spin-orbit interaction is stronger, due to a larger $p$ and a smaller $x$.}
\label{fig:GaMnAs_comp2}
\end{center}
\end{figure}

Analogous considerations apply for Figs. (~\ref{fig:GaMnAs_comp1}) and (~\ref{fig:GaMnAs_comp2}), which show results for the 
four-band model related to (Ga,Mn)As. These figures show results similar to those obtained in the 
strong spin-orbit limit of the M2DEG (Fig. ~\ref{fig:M2DEG_comp2}). 
Overall, our study indicates that the \emph{torque-correlation} formula captures the intra-band contributions accurately when the vertex corrections are unimportant, while it is less reliable for inter-band contributions unless the predominant inter-band transitions connect states that are close in energy. 
The torque-correlation formula has the practical advantage that it correctly gives a zero spin relaxation rate when there is 
no spin-orbit coupling in the band structure and spin-independent disorder.  The damping it captures derives entirely from
spin-orbit coupling in the bands.  It therefore incorrectly predicts, for example, that the damping rate vanishes when
spin-orbit coupling is absent in the bands and the disorder potential is spin-dependent.  Nevertheless, assuming that 
the dominant disorder is normally spin-independent, the $K$-formula may have a pragmatic edge over the $s^{x}$-formula 
in weakly spin-orbit coupled systems.  In strongly spin-orbit coupled systems there appears to be little advantage of 
one formula over the other.  We recommend that inter-band and intra-band contributions be evaluated separately when $\alpha_{G}$ is evaluated using the torque-correlation formula.  For the intra-band contribution the 
$s^{x}$ and $K$ life-time formulas are identical.  The model calculations reported here 
suggest that vertex corrections to the intra-band contribution do not normally have an overwhelming importance.
We conclude that $\alpha_{G}$ can be evaluated relatively reliably when the intra-band contribution dominates.
When the inter-band contribution dominates it is important to assess whether or not the dominant contributions are 
coming from bands that are nearby in momentum space, or equivalently whether or not the matrix elements which 
contribute originate from pairs of bands that are energetically spaced by much less than the exchange spin-splitting
at the same wavevector.  If the dominant contributions are from nearby bands, the damping estimate should have the 
same reliability as the intra-band contribution.  If not, we conclude that the $\alpha_{G}$ estimate should be
regarded with caution.

To summarize, this article describes an evaluation of Gilbert damping for two simple models, a two-dimensional electron-gas ferromagnet model with Rashba spin-orbit interactions and a four-band model which provides an approximate description of (III, Mn)V of ferromagnetic semiconductors. Our results are exact in the sense that they combine time-dependent mean field theory\cite{alphaI} with an impurity ladder-sum to all orders, hence giving us leverage to make the following statements.  First, previously neglected higher order vertex corrections become quantitatively significant when the intrinsic spin-orbit interaction is larger than the exchange splitting. Second, strong intrinsic spin-orbit interaction leads to the the supremacy of intra-band contributions in (Ga,Mn)As, with the corresponding  monotonic decay of the Gilbert damping as a function of disorder.  Third, the spin-torque formalism used in \emph{ab-initio} calculations of the Gilbert damping is quantitatively reliable as long as the intra-band contributions dominate \emph{and} the exchange field is weaker than the spin-orbit splitting; if these conditions are not met, the use of the spin-torque matrix element in a life-time approximation 
formula offers no significant improvement over the original spin-flip matrix element.

\acknowledgements
The authors thank Keith Gilmore and Mark Stiles for helpful discussions and feedback.  This work was supported by the
Welch Foundation and by the National Science Foundation under grant DMR-0606489.


\begin{thebibliography}{30}
\bibitem{disorder} This is not the most general type of disorder for quasiparticles with spin $>$ 1/2, but it will be sufficient for the purpose of this work.
\bibitem{caveat1} We assume that the spins of  magnetic impurities are \emph{frozen} along the \emph{static} part of the exchange field. In reality, the direction of the impurity spins is a dynamical variable that is influenced by the magnetization precession.
\bibitem{B and C} $C(1)$ and $C(\infty)$ differ by as much as 25\%; the disparity between $B(1)$ and $B(\infty)$ may be even larger.
\bibitem{greens} The disorder dependence in $\alpha_{G}$ originates not only from the vertex part, but from the Green's functions as well. It is useful to recall that $\int G_{\sigma} G_{-\sigma} \propto 1/(b+i\text{sg}(\sigma)\gamma)$ and $\int G_{\sigma} G_{\sigma} \propto 1/\gamma$.
\bibitem{weak SO} Notwithstanding that the four-band model is a SO $\to\infty$  limit of the more general six-band model, we shall tune the effective spin-orbit strength via $p$ (hole concentration) and $\gamma_{3}$.
\bibitem{caveat2} Strictly speaking,  it is $|s^{x}_{\alpha,\beta}|^{2}\simeq |K_{\alpha,\beta}|^{2}$ what is needed, rather than 
$s^{x}_{\alpha,\beta}\simeq K_{\alpha,\beta}$. The former condition is less demanding, and can occasionally be satisfied when $E_{\alpha}-E_{\beta}$ is of the order of the exchange splitting.
\end{thebibliography}
\end{document}